\documentclass[prd,superscriptaddress,nofootinbib,amsmath,amssymb,aps,11pt]{revtex4-2}

\usepackage{bm}
\usepackage{amsfonts}
\usepackage{latexsym}
\usepackage[utf8]{inputenc}
\usepackage{graphicx}
\usepackage{amsmath}
\usepackage{palatino}
\usepackage{mathpazo}
\usepackage[normalem]{ulem}
\usepackage{xcolor}

\usepackage{booktabs}
\usepackage{dcolumn}

\usepackage{natbib}

\begin{document}
\title{Neutron stars in Gauss-Bonnet gravity -- nonlinear scalarization and gravitational phase transitions}
\author{Daniela D. Doneva}
\email{daniela.doneva@uni-tuebingen.de}
\affiliation{Theoretical Astrophysics, IAAT, University of T\"ubingen, 72076 T\"ubingen, Germany}
\affiliation{INRNE - Bulgarian Academy of Sciences, 1784  Sofia, Bulgaria}

\author{Christian J. Kr\"uger}
\email{christian.krueger@tat.uni-tuebingen.de}
\affiliation{Theoretical Astrophysics, IAAT, University of T\"ubingen, 72076 T\"ubingen, Germany}

\author{Kalin V. Staykov}
\email{kstaykov@phys.uni-sofia.bg}
\affiliation{Department of Theoretical Physics, Faculty of Physics, Sofia University, Sofia 1164, Bulgaria}

\author{Petar Y. Yordanov}
\email{pyordanov@phys.uni-sofia.bg}
\affiliation{Department of Theoretical Physics, Faculty of Physics, Sofia University, Sofia 1164, Bulgaria}

\date{April 2023}

\begin{abstract}
    It was recently discovered that scalarized neutron stars in scalar-tensor theories can undergo a gravitational phase transition to a non-scalarized (GR) state. Surprisingly, even though the driving mechanism is totally different, the process resembles closely the first-order matter phase transition from confined nuclear matter to deconfined quark matter in neutron star cores. The studies until now were limited, though, to only one theory of gravity and a limited range of parameters. With the present paper, we aim at demonstrating that gravitational phase transitions are more common than expected. More specifically, we show that the phenomenon of nonlinear scalarization is present for neutron stars in Gauss-Bonnet gravity leading to the possibility of gravitational phase transition. Moreover, it can be observed for a wide range of parameters so no fine-tuning is needed. This solidifies the conjecture that gravitational phase transitions are an important phenomenon for compact objects and their astrophysical implications deserve an in-depth study.
\end{abstract}

\maketitle

\section{Introduction}
Electromagnetic and gravitational wave observations have been developing rapidly in the last decade. Several new instruments are being built or planned that will finally reach the capabilities to test fundamental physics in fine detail. Among the most interesting astrophysical objects in this context are neutron stars that offer an example of physics at its extremes. The density in their cores is larger than what we can produce with terrestrial experiments that allows us to explore the matter equation of state through astrophysical observations. On the other hand, the compactness of neutron stars is so large that strong gravity effects start being important. Thus, a modification of general relativity (GR) will eventually lead to differences in the neutron star structure and dynamics. For that reason, a lot of efforts in the past decades were devoted to building neutron star models in generalizations of Einstein's theory and predicting their astrophysical manifestations (see e.g. \cite{Berti:2015itd, Doneva:2017jop} for a review). 

It is often not possible, though, to test independently the neutron star equation of state and the theory of gravity. The reason is the well-known degeneracy between effects related to the equation of state uncertainty and to modifications of general relativity \cite{Shao:2019gjj}. From first sight, this fact speaks against using neutron stars to test gravity, and often black holes are preferred as ``cleaner'' objects. One should keep in mind, though, that neutron stars have much richer phenomenology with a plethora of observations. That is why, despite the obstacles, they still provide the best constraints in a number of modified theories of gravity \cite{LIGOScientific:2018dkp,Zhao:2022vig,Silva:2020acr}. 

It was recently discovered that neutron stars in scalar-tensor theories of gravity can undergo gravitational phase transitions from a state with nonzero scalar hair to a GR state, thus emitting completely their scalar field \cite{Kuan:2022oxs}. Such a process happens when we have two disconnected sequences of stable solutions (scalarized and non-scalarized ones) and a transition between them is possible via some astrophysical process, e.g. accretion, slowdown of a post-merger neutron star, etc. This process has striking similarities with the matter first-order phase transition from confined nuclear matter to deconfined quarks in pure GR (the so-called twins)  \cite{kam81,Glendenning:1998ag,sch00,shaf02}. Such twin stars in GR offer a plethora of very interesting astrophysical implications and signatures that can be traced in observational data \cite{Most:2018eaw,Bauswein:2018bma,Weih:2019xvw,Liebling:2020dhf,Most:2019onn,Blacker:2020nlq} (for a review see e.g. \cite{Blaschke:2019tbh}).  Thus, the results in \cite{Kuan:2022oxs} demonstrate that modified gravity can mimic specifics of the high-density equation of state on a much more sophisticated level than expected.

The neutron star model in \cite{Kuan:2022oxs} assumes as an underlying theory of gravity the Damour-Esposito-Farese scalar-tensor theory \cite{dam93} with a massive scalar field \cite{Ramazanoglu16,Yazadjiev16}. The gravitational phase transitions in this case happen generally for larger neutron star masses and resemble the high-density twin star branches considered e.g. in \cite{Christian:2019qer}. It was unclear, though, whether the gravitational phase transitions  discovered in \cite{Kuan:2022oxs} are an isolated phenomenon occurring in a specific modified theory of gravity  in a relatively narrow parameter range. That is why searching for other gravitational theories which possess such a property is an important development of the problem. Moreover, the threshold density where first-order matter phase transition happens can vary significantly \cite{Benic:2014jia}, something that is not so easily controllable at least for the model in \cite{Kuan:2022oxs}.

The present paper aims to give an answer to these open questions. More specifically we consider another modified theory of gravity, namely scalar-Gauss-Bonnet (henceforth sGB) theory. As it turns out, sequences of neutron stars exist that resemble even more closely the standard picture of twin stars in general relativity. More specifically, the origin of the scalarized branch can be at an arbitrarily small central energy density and the radius of the scalarized phases decreases with respect to GR. What differs from the standard picture of neutron star scalarization in sGB gravity \cite{Silva:2017uqg, Doneva:2017duq, Xu:2021kfh} (where such an effect is not observed) is the choice of a more general form of the coupling function between the Gauss-Bonnet invariant and the scalar field. Such a modification was inspired by the black hole phase transitions and the related effect of nonlinear scalarization \cite{Doneva:2021tvn}. In that case, two disconnected branches of black hole solutions exist -- one scalarized and one with a zero scalar field (Schwarzschild solution), that can have interesting astrophysical implications \cite{Doneva:2022byd}.  

The paper is organized as follows. Section II  is devoted to the theoretical setup of the problem, reviewing very briefly the topic of generating neutron star solutions in sGB gravity. The following Section III focuses on the numerical construction of the desired solutions admitting gravitational phase transitions. After we present the results in Section~\ref{sec:results}, the paper finishes with a discussion.

\section{scalar-Gauss-Bonnet gravity}

In the present paper we consider the  sGB gravity in the presence of  matter. The most general form of the action in this case has the following form:  
\begin{eqnarray}
S=&&\frac{1}{16\pi}\int d^4x \sqrt{-g} 
\Big[R - 2\nabla_\mu \varphi \nabla^\mu \varphi  - V(\varphi)
+ \lambda^2 f(\varphi){\cal R}^2_{GB} \Big] + S_{\rm matter}  (g_{\mu\nu},\chi) ,\label{eq:quadratic}
\end{eqnarray}
where $R$ denotes the Ricci scalar with respect to the spacetime metric $g_{\mu\nu}$ and $\nabla_{\mu}$ is the covariant derivative with respect to $g_{\mu\nu}$. $V(\varphi)$ is the potential and  $f(\varphi)$ is the coupling function for the scalar field $\varphi$.  The Gauss-Bonnet coupling constant $\lambda$ has  dimension of $length$ and ${\cal R}^2_{GB}$ denotes the Gauss-Bonnet invariant defined by ${\cal R}^2_{GB}=R^2 - 4 R_{\mu\nu} R^{\mu\nu} + R_{\mu\nu\alpha\beta}R^{\mu\nu\alpha\beta}$ where $R_{\mu\nu}$ and $R_{\mu\nu\alpha\beta}$ are  the Ricci tensor and  the Riemann tensor respectively. $S_{\rm matter}$ is the matter action where the matter fields are collectively denoted by $\chi$. 

The field equations derived from the action (\ref{eq:quadratic}) are 
\begin{eqnarray}\label{FE}
&&R_{\mu\nu}- \frac{1}{2}R g_{\mu\nu} + \Gamma_{\mu\nu}= 2\nabla_\mu\varphi\nabla_\nu\varphi -  g_{\mu\nu} \nabla_\alpha\varphi \nabla^\alpha\varphi -\frac{1}{2}g_{\mu\nu}V(\varphi)  + 8\pi T^{\rm matter}_{\mu\nu},\\
\label{eq:fe_scalar}
&&\nabla_\alpha\nabla^\alpha\varphi= \frac{1}{4}\frac{dV(\varphi)}{d\varphi}  -  \frac{\lambda^2}{4} \frac{df(\varphi)}{d\varphi} {\cal R}^2_{GB},
\end{eqnarray}
where  $T^{\rm matter}_{\mu\nu}$ is the matter energy momentum tensor.  $\Gamma_{\mu\nu}$ is defined by 
\begin{eqnarray}
\Gamma_{\mu\nu}&=& - R(\nabla_\mu\Psi_{\nu} + \nabla_\nu\Psi_{\mu} ) - 4\nabla^\alpha\Psi_{\alpha}\left(R_{\mu\nu} - \frac{1}{2}R g_{\mu\nu}\right) + 
4R_{\mu\alpha}\nabla^\alpha\Psi_{\nu} + 4R_{\nu\alpha}\nabla^\alpha\Psi_{\mu} \nonumber \\ 
&& - 4 g_{\mu\nu} R^{\alpha\beta}\nabla_\alpha\Psi_{\beta} 
+ \,  4 R^{\beta}_{\;\mu\alpha\nu}\nabla^\alpha\Psi_{\beta} 
\end{eqnarray}  
with 
\begin{eqnarray}
\Psi_{\mu}= \lambda^2 \frac{df(\varphi)}{d\varphi}\nabla_\mu\varphi .
\end{eqnarray}
In the present study we will concentrate on the theory with a vanishing potential as a case study (i.e., we set $V(\varphi) = 0$).

In addition to the field equations, in order for a neutron star model to be constructed,  we have to add the equation for hydrostatic equilibrium of the fluid which can be derived  from the condition
\begin{eqnarray}\label{BID}
\nabla^{\mu}T^{\rm matter}_{\mu\nu}=0. 
\end{eqnarray}

For the purpose of the present paper (with exception of the nonlinear saturation study in Sec.~\ref{sec:nonlinear_saturation}) we adopt the standard ansatz for  static and spherically  symmetric metric  
\begin{eqnarray}
\label{eq:metric}
ds^2= - e^{2\Phi(r)}dt^2 + e^{2\Lambda(r)} dr^2 + r^2 (d\theta^2 + \sin^2\theta d\phi^2 ).
\end{eqnarray}   

We choose the matter source to be a perfect fluid with $T^{\rm matter}_{\mu\nu}=(\rho + p)u_{\mu}u_{\nu} + pg_{\mu\nu}$ 
where $\rho$, $p$ and $u^{\mu}$ are the energy density, pressure and 4-velocity of the fluid, respectively. We also require the perfect  
fluid and the scalar field to a have static and spherically symmetric distribution. For the explicit form of the dimensionally reduced field equations and the equation for hydrostatic equilibrium, we refer the interested reader to \cite{Doneva:2017duq}.  In order for those equations to be solved, they should be supplemented with an equation of state (EoS) and proper boundary conditions. The EoS will be commented on in the following section. Concerning the boundary conditions, they are the natural ones -- regularity at the center of the star, 

\begin{equation}
	\left.\Lambda\right|_{r\rightarrow 0}\rightarrow 0,\quad \left. \frac{d\Phi}{dr}\right|_{r\rightarrow 0}\rightarrow 0, \quad \left.\frac{d\varphi}{dr}\right|_{r\rightarrow 0}\rightarrow 0
\end{equation}
and asymptotic flatness at spacial infinity,
\begin{equation}
	\left.\Lambda\right|_{r\rightarrow \infty}\rightarrow 0,\quad \left.\Phi\right|_{r\rightarrow \infty}\rightarrow 0, \quad \left.\varphi\right|_{r\rightarrow \infty}\rightarrow 0.
\end{equation}

As it is explained in detail in \cite{Doneva:2017duq}, the regularity at the stellar center can not be fulfilled typically at large central energy densities. Therefore, the scalarized branches of solutions are terminated at a fixed $\rho$. 
The regularity condition can be derived in the following manner.  The metric functions and the scalar field are substituted with their series expansions
\begin{equation}
\Lambda=\Lambda_0 + \Lambda_1 r+\frac{1}{2} \Lambda_2 r^2  + O(r^3)\;\;\;\Phi=\Phi_0+\Phi_1 r+\frac{1}{2} \Phi_2 r^2 + O(r^3) \;\;\; \varphi=\varphi_0+\varphi_1 r + \frac{1}{2} \varphi_2 r^2 + O(r^3)
\end{equation}
in the explicit form of the field equations. Comparing the coefficients of different powers of $r$ then leads to constraints on the Taylor coefficients of the series expansions.

Following this path, it is straightforward to conclude that $\Lambda_0=0$, $\Lambda_1=0$, $\Phi_1=0$, and $\varphi_1=0$ (in line with the regularity of the solution at the center of the star). In addition, the following condition is also obtained
\begin{equation}\label{eq:SolutionConstrain}
9  \lambda^4  \left(\frac{df}{d\varphi}(\varphi_0)\right)^2 (\Lambda_2)^4 - 72 \pi \lambda^4 p_0  \left(\frac{df}{d\varphi}(\varphi_0)\right)^2   (\Lambda_2)^3 - 6\pi \rho_0  \Lambda_2 + 16 \pi^2 \rho_{0}^2=0.
\end{equation}
This is a fourth-order algebraic equation for $\Lambda_2$ that depends on the values of the pressure $p_0$ and energy density $\rho_0$ at the center of the star, as well as the central value of the scalar field $\varphi_0$. In case no real root and thus a solution for $\Lambda_2$ exists, regular scalarized neutron stars can not be constructed.

\section{Nonlinear scalarization and scalar field coupling function}

Our goal is to extend the previous studies of neutron stars in sGB gravity \cite{Doneva:2017duq}  beyond the standard curvature-induced spontaneous scalarization  and explore an effect analogous to the nonlinear scalarization observed for sGB black holes \cite{Doneva:2021tvn}. We are mostly concentrated on the scalarization effect, hence only one equation of state is employed, namely APR4 \cite{Akmal:1998cf} with its piecewise polytropic approximation \cite{Read:2008iy}.

Let us first comment in more detail on the different types of scalarization. The key property of spontaneous scalarization is that the GR solutions are also solutions in the specific modified theory of gravity. For strong enough curvature, though, the zero scalar field neutron stars can lose stability giving rise to a scalarized branch of solutions. This phenomenon was first discovered for neutron stars in the famous Damour-Esposito-Farese (DEF) model \cite{Damour:1993hw}. A similar effect is observed for charged black holes in scalar-tensor theories described by nonlinear electrodynamics \cite{Stefanov:2007eq}. In both cases, the scalar field is sourced by the matter having nonzero trace of the energy-momentum tensor. The picture differs, though, for black hole or neutron star scalarization in sGB gravity where the source of the scalar field is the curvature itself through the direct coupling of the scalar field to the Gauss-Bonnet invariant (see Eq. \eqref{eq:quadratic}). This is the so-called curvature-induced scalarization  \cite{Doneva:2017bvd,Silva:2017uqg,Antoniou:2017acq}.

In order for the spontaneous scalarization to realize one should choose a coupling function $f(\varphi)$ having zero first derivative with respect to $\varphi$ at $\varphi=0$, i.e. $df(\varphi)/d\varphi|_{\varphi=0}=0$. In that way, $\varphi=0$  is always a solution of the field equations \eqref{FE}. Let us now consider a perturbation of the scalar field equation Eq.~\eqref{FE} on top of the GR background (where we assume $V(\varphi)=0$ for simplicity),
\begin{equation}
    \left( \nabla_\alpha\nabla^\alpha - m^2_{\rm eff} \right)\delta \varphi= 0\,.
\end{equation}
Here the scalar field effective mass is defined as
\begin{equation}
    m^2_{\rm eff} = - \frac{\lambda^2}{4}\left.\frac{d^2 f(\varphi)}{d\varphi^2 }\right|_{\varphi=0} {\cal R}^2_{GB}\,.
\end{equation}
While for non-rotating black holes always ${\cal R}^2_{GB}>0$, static neutron stars have regions of both positive and negative ${\cal R}^2_{GB}$. Therefore, when $d^2f(\varphi)/d\varphi^2|_{\varphi=0}\ne 0$ the square of the scalar field effective mass $m^2_{\rm eff}$ can become negative which ignites the scalar field development (see \cite{Doneva:2022ewd} for a detailed review on the topic).

Clearly, the second derivative of the coupling function $d^2f(\varphi)/d\varphi^2$ is responsible for the destabilization of a GR compact object. If unstable, any small perturbation will ignite the scalar field leading to a scalarized solution. For $d^2f(\varphi)/d\varphi^2=0$ the effective mass $m^2_{\rm eff}=0$ and the GR solutions are always linearly stable. It was recently shown, though, that black holes endowed with scalar hair can still be present in that case, being stable and coexisting with the linearly stable Schwarzschild black holes \cite{Doneva:2021tvn,Blazquez-Salcedo:2022omw} (see also \cite{Zhang:2023jei,Lai:2023gwe,Jiang:2023yyn}). The growth of the scalar field, in that case, can not be ignited by a small arbitrary perturbation. A larger nonlinear perturbation, on the other hand, can lead to scalar field development since the hairy black holes are in general energetically favored over the GR solutions. That is why this phenomenon was dubbed nonlinear scalarization. The transition between a Schwarzschild black hole and a scalarized one happens with a jump that can have interesting astrophysical manifestations. A similar picture was also observed for black holes in Einstein-Maxwell-dilaton gravity \cite{Blazquez-Salcedo:2020nhs,LuisBlazquez-Salcedo:2020rqp}. In the following section, we will demonstrate a similar phenomenon, however, in the present case for neutron stars in sGB gravity. This will eventually be connected with the gravitational phase transitions.

The simplest coupling function that satisfied the criteria for standard spontaneous scalarization with $m^2_{\rm eff}<0$ is $f(\varphi)=\varphi^2$. The nonlinear scalarization is realized when the leading order term is proportional to a higher order of $\varphi$, e.g. $f(\varphi)=\varphi^4$ \cite{Doneva:2021tvn,Blazquez-Salcedo:2020nhs}. A combination of both terms can lead to a mixture of spontaneous and nonlinear scalarization at least for black holes \cite{Doneva:2021tvn}. Namely, for a function of the type $f(\varphi)=\varphi^2 + a\varphi^4$, where $a$ is a constant, there is a range of black hole masses where the Schwarzschild solution is linearly unstable and one can scalarize a black hole by a small scalar field perturbation. For somewhat larger masses, it is possible to have a co-existence of linearly stable Schwarzschild black holes and stable scalarized solutions \cite{Doneva:2021tvn,Blazquez-Salcedo:2022omw,Silva:2018qhn,Minamitsuji:2018xde}.

Our goal in the present paper is to translate these results to neutron stars. We exploit two scalar field coupling functions designated as $f_1(\varphi)$ and $f_2(\varphi)$ respectively, both containing quadratic and quartic terms 
\begin{equation}\label{eq:coupling_f1}
    f_1(\varphi) = -\frac{1}{2\beta}(1- e^{-\beta(\varphi^2 + \kappa\varphi^4)}),
\end{equation}
\begin{equation}\label{eq:coupling_f2}
    f_2(\varphi) = \frac{1}{2\beta}(1- e^{-\beta(\varphi^2 + \kappa\varphi^4)}).
\end{equation}

Clearly, the two couplings differ only in their sign. The equilibrium neutron stars built with either Eq.~\eqref{eq:coupling_f1} or Eq.~\eqref{eq:coupling_f2} show very different scalar field profiles, though, as detailed in \cite{Doneva:2017duq}. In addition, we have chosen an exponential form of the coupling functions instead of a simple polynomial, just for numerical convenience. Such a choice makes also the comparison with previous studies on black holes easier \cite{Doneva:2017duq}. Note that only one of these functions can lead to black hole scalarization in the static case, namely Eq.~\eqref{eq:coupling_f2}.

In accordance with the discussion above, both $f_1(\varphi)$ and $f_2(\varphi)$  exhibit the standard neutron star curvature-induced spontaneous scalarization when $\kappa = 0$. The introduction of a nonzero $\kappa$, though, will be the trigger of the nonlinear scalarization for a certain range of central energy densities. 

\section{Results}
\label{sec:results}

For both coupling functions $f_1$ and $f_2$ we fix two pairs of values for the parameters $(\lambda,\beta)$ and study a wide range of $\kappa$ starting from $\kappa=0$ and reaching the regime where nonlinear scalarized phases and gravitational phase transitions could appear. As evident from \cite{Danchev:2021tew,Kuan:2023trn}, the considered $(\lambda,\beta)$ are already excluded from binary pulsar observations in the case of a massless scalar field. This is a well-known limitation for scalarized neutron star models also in other theories of gravity (see e.g. \cite{Doneva:2022ewd}). If one considers a nonzero scalar field mass these constraints can be easily evaded. In that case, the scalar field is exponentially suppressed at a length scale larger than its Compton wavelength thus leading to a zero scalar charge and zero scalar radiation.

%-----------------------
The rule of thumb is that in order to evade the binary pulsar constraints, the Compton wavelength should be much smaller compared to the orbital separation of a binary. A scalar field mass, as small as $m_\varphi \sim 10^{-16}$ would be enough for that purpose \cite{Ramazanoglu16,Yazadjiev16} and practically no constraints can be put on $(\lambda,\beta)$ from the binary pulsars. Very importantly, such a small mass has practically no influence on the neutron star properties and dynamics compared to the massless case \cite{Kuan:2023trn,Xu:2021kfh}. This justifies the fact that the models presented below are calculated in the massless case, i.e. with zero scalar field potential. Similar approaches of including a scalar field mass when studying astrophysical implications are often employed in other modified theories, such as scalar-tensor theories of gravity and DEF models in particular (see e.g. \cite{Sperhake:2017itk}).

In the numerical results presented below, we use the dimensionless parameter
\begin{eqnarray}
\lambda \to \frac{\lambda}{R_0}
\end{eqnarray}
where $R_0 = 1.47664$ km corresponds to one solar mass.

The scalar charge $D$, presented below, is defined by the series expansion of the scalar field at spacial infinity
\begin{equation}
	\varphi \simeq \frac{D}{r} + O(1/r^2).
\end{equation}
It is also a dimensional quantity and we present it in units of $R_0$. Clearly, $D=0$ in the massive scalar field case because the scalar field decays exponentially at large distances. Nevertheless, we decided to keep this characteristic in our presentation (calculated assuming $m_\varphi=0$) since it is instructive and can be used as a comparison with previous results on scalarized neutron stars, including the DEF models.

\subsection{Coupling function $f_1(\varphi)$}

\begin{figure}[]
	\includegraphics[width=0.45\textwidth]{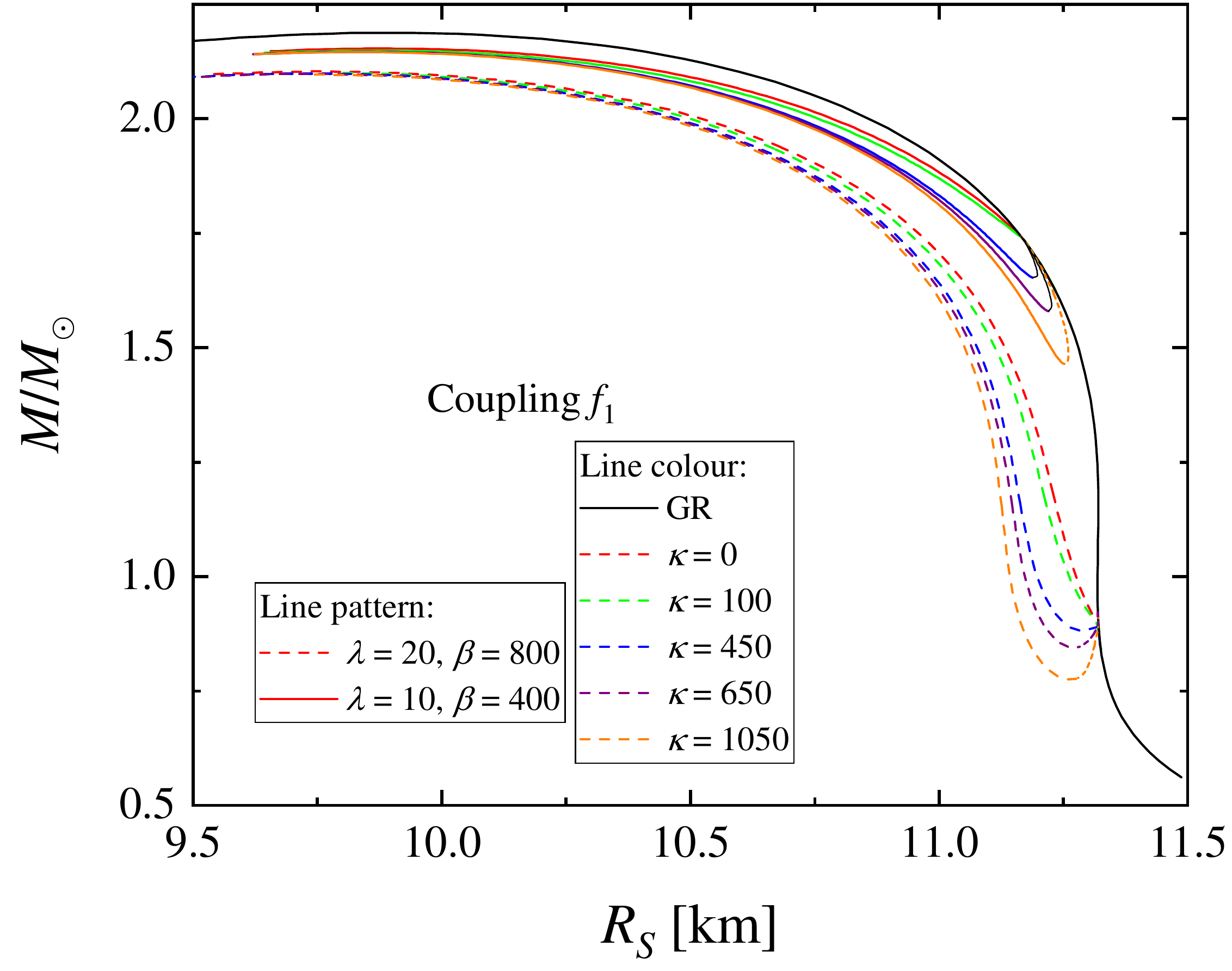}
	\includegraphics[width=0.45\textwidth]{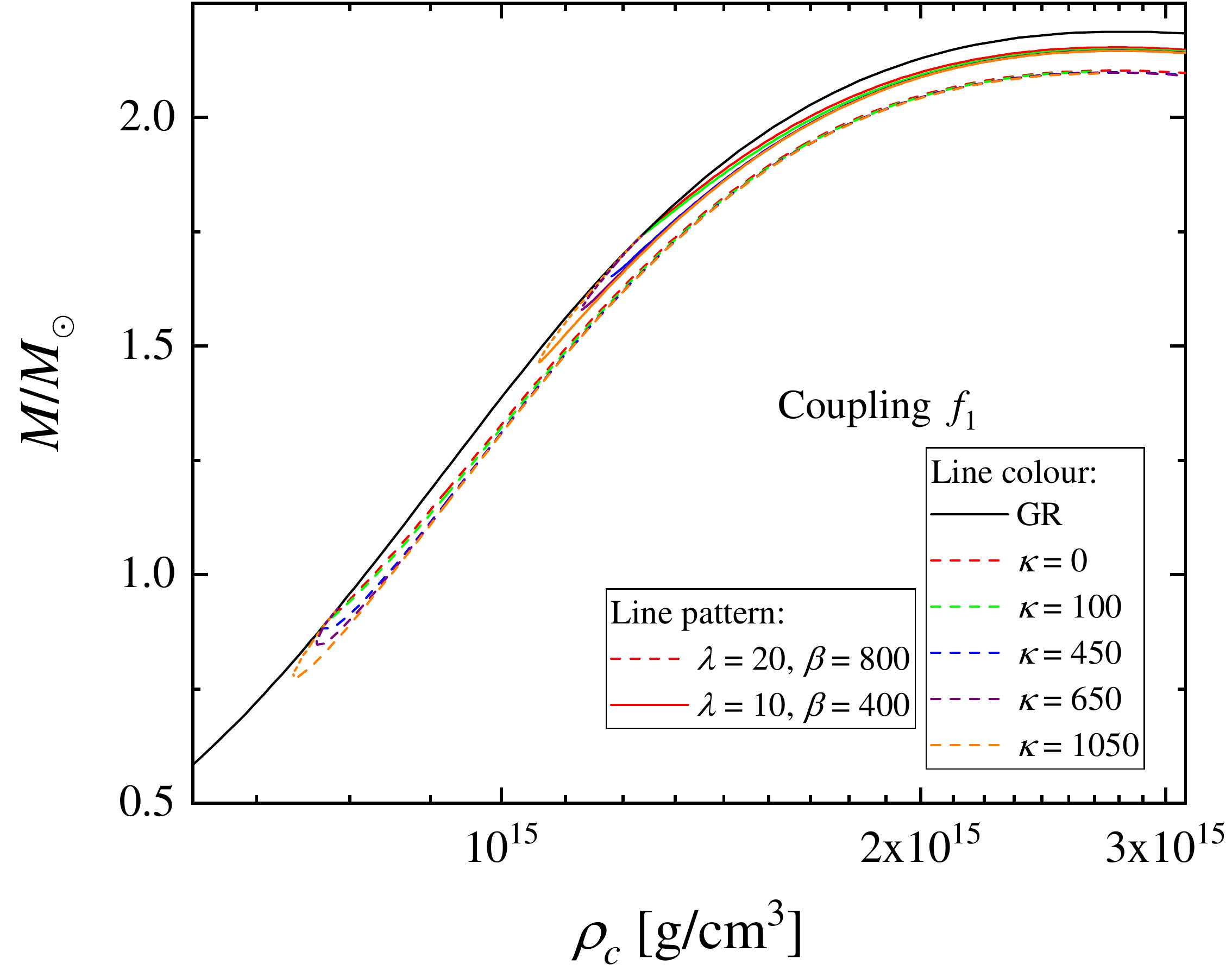}
	\includegraphics[width=0.45\textwidth]{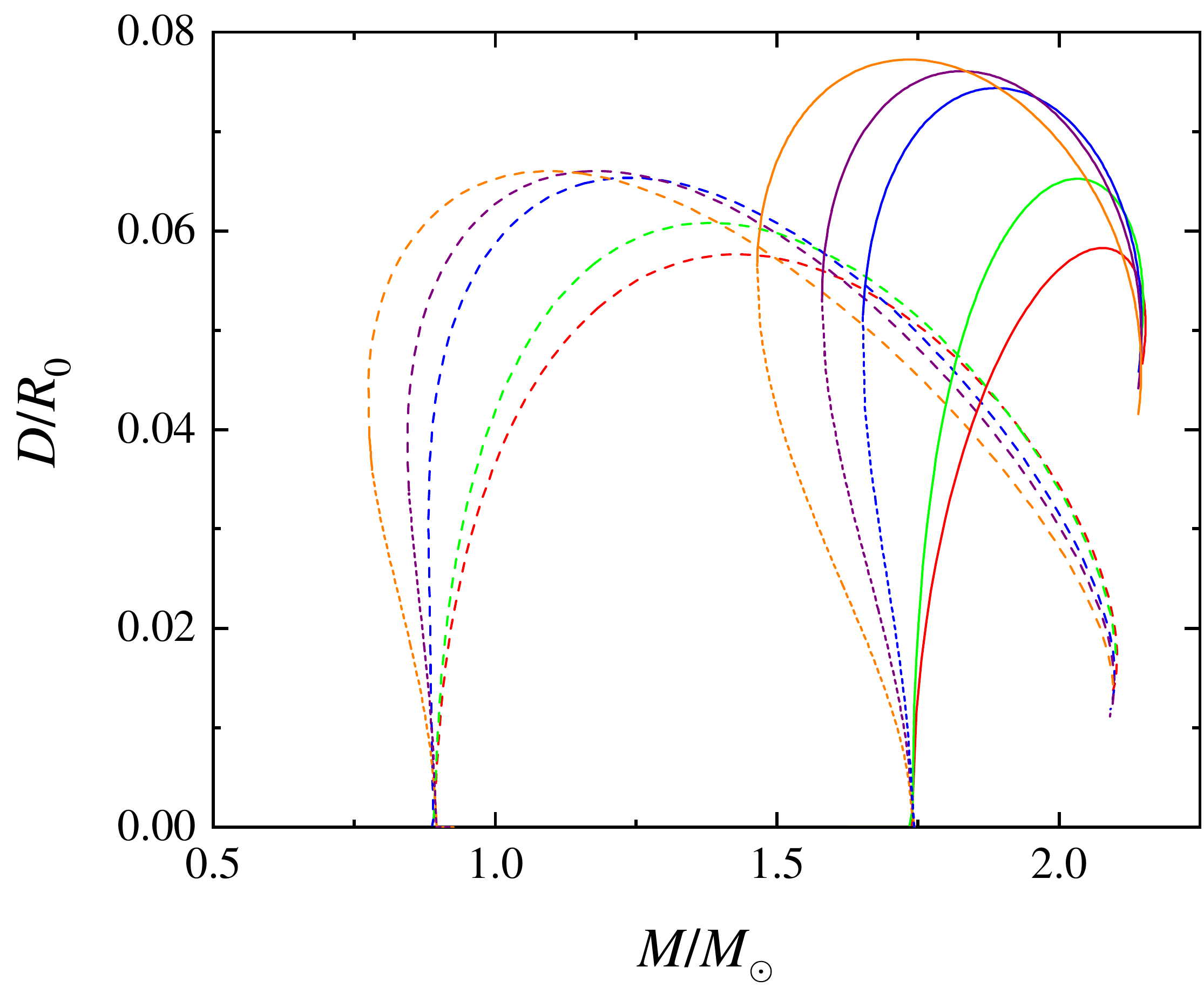}
	\caption{ \textit{(top, left)} Neutron star mass as a function of its radius for several choices of the coupling parameters $\lambda,\beta$ and $\kappa$. The different pairs $(\lambda,\beta)$ are in depicted with different line patterns, while different values of $\kappa$ are shown in different colours. \textit{(top, right)} Neutron star mass as a function of its central energy density for the same sequences of solution. \textit{(bottom)} The  scalar charge ,in units of $R_0$, as a function of the mass of the star. The presented results in these graphs are for the coupling function $f_1(\varphi)$.}
	\label{fig:M_c1}
\end{figure}

We start our study with the coupling function $f_1(\varphi)$. Let us remind the reader that this is the case when black hole scalarization is not possible and only neutron stars can develop a nontrivial scalar field. Two pairs of $(\lambda,\beta)$ are investigated: $(\lambda,\beta) = (10,400)$ and $(\lambda,\beta) = (20,800)$. In the former case, the bifurcation point where scalarized solutions originate from the GR ones is located close to the maximal neutron star mass. The latter combination of parameters results in a bifurcation at a relatively small neutron star mass. The value of $\beta$ is kept large enough in both cases so that the deviation from GR is moderate. 

In the upper panels of Fig.~\ref{fig:M_c1}, we plot the mass of the star as a function either of its radius (top-left panel) or as a function of the central energy density (top-right panel). The case of standard scalarization in sGB gravity reported in \cite{Doneva:2017duq} corresponds to $\kappa = 0$. In both plots, the sequences of solutions remain quantitatively very similar to the GR case for small $\kappa$, and only the magnitude of the deviation from GR increases. This changes with increasing $\kappa$, though. Namely, for large $\kappa$ two distinct parts can be distinguished in the sequences. The first branch starts from the bifurcation point and spans until the minimum mass of the scalarized sequence. As we will see below, we have very strong indications that it is unstable. That is why we have depicted it with dotted lines in each of the figures. We would like to remind the reader that the GR branch is also unstable above the bifurcation point for a given $\lambda$. The minimum mass for such a sequence of solutions depends strongly on the parameter $\kappa$ (for fixed $(\lambda,\beta)$) -- the higher the value of $\kappa$ is, the lower this minimal mass is. At that minimal mass, the going downward branch transitions smoothly into a second scalarized branch, going upwards. Depending on the rest of the parameters, the second scalarized branch could reach maximal mass or stop before that due to violation of the regularity condition (\ref{eq:SolutionConstrain}). Based on the analysis below, we believe that it is most probably stable.

The observed behavior resembles the nonlinear scalarization for black holes \cite{Doneva:2021tvn}. Namely, there is an interesting region starting from the minimum central energy density where scalarized solutions exist and reaching up to the bifurcation point. There, both a potentially stable scalarized and the linearly stable GR branches co-exist \cite{Blazquez-Salcedo:2022omw}. No smooth connection (in the form of a sequence of stable solutions) between them is present and thus transitioning will happen with a jump. Similar branch structure lies also in the foundation of the gravitational phase transitions discovered in the DEF models \cite{Kuan:2022oxs}. In sGB gravity it is much more controllable, though---the phase transition can happen at arbitrarily small/large central energy densities dependent on $\lambda$. This is in contrast to the DEF models where such transitions could appear only close to the maximum neutron star mass. Another difference between the two theories is that the ``gap'' between the two potentially stable branches (the scalarized one and the GR branch) can be considerably larger in sGB gravity.

In the bottom panel  of Fig. \ref{fig:M_c1}, for completeness, we present the  scalar charge, in units of $R_0$, as a function of the mass of the star. Similar to the DEF model it starts from zero at the bifurcation point and after reaching a maximum it decreases. The scalar charge never reaches $D=0$ again, though, due to violation of the regularity condition. Thus, a second scalarization point does not exist in contrast with the DEF model.

\begin{figure}[]
	\includegraphics[width=0.49\textwidth]{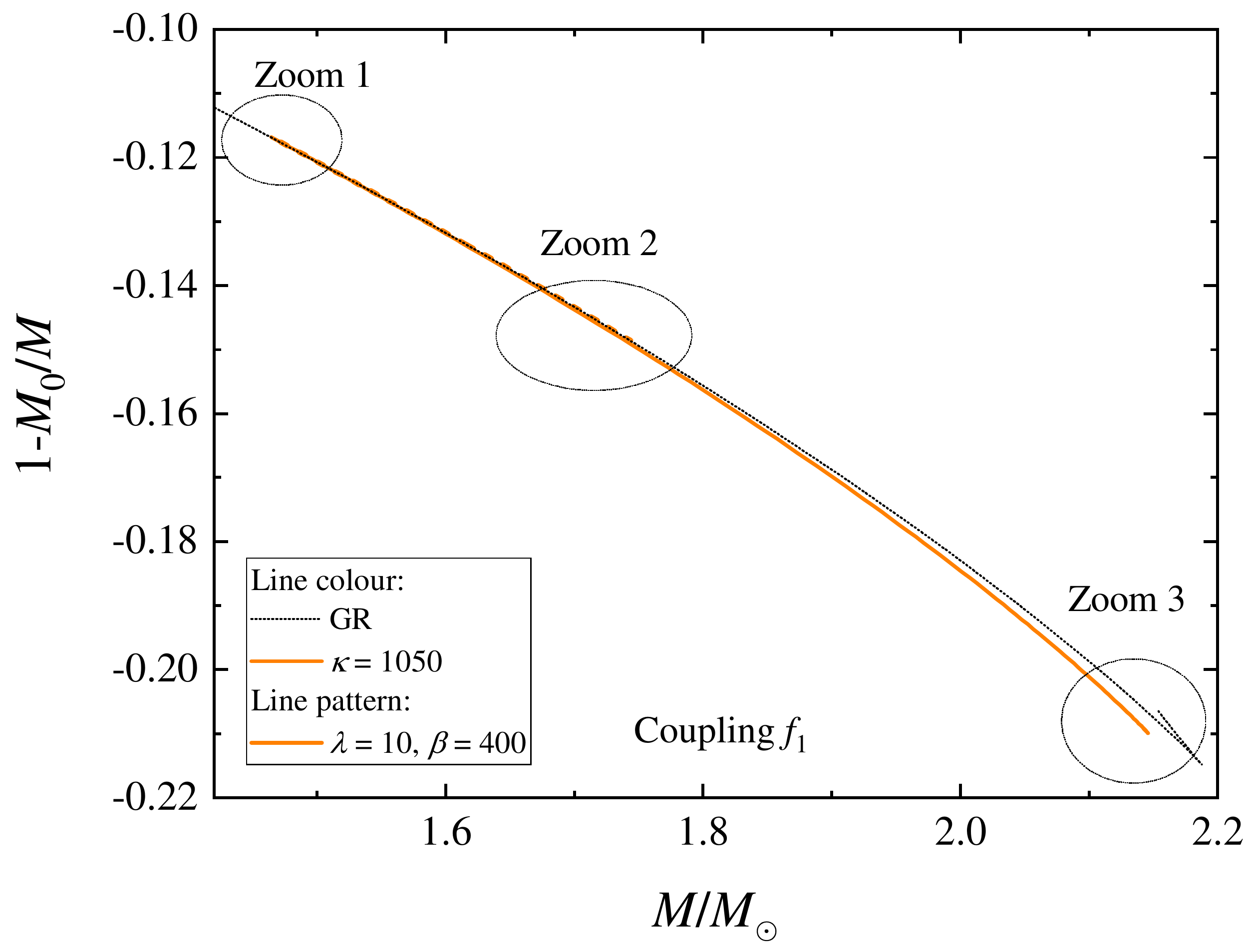}
	\includegraphics[width=0.49\textwidth]{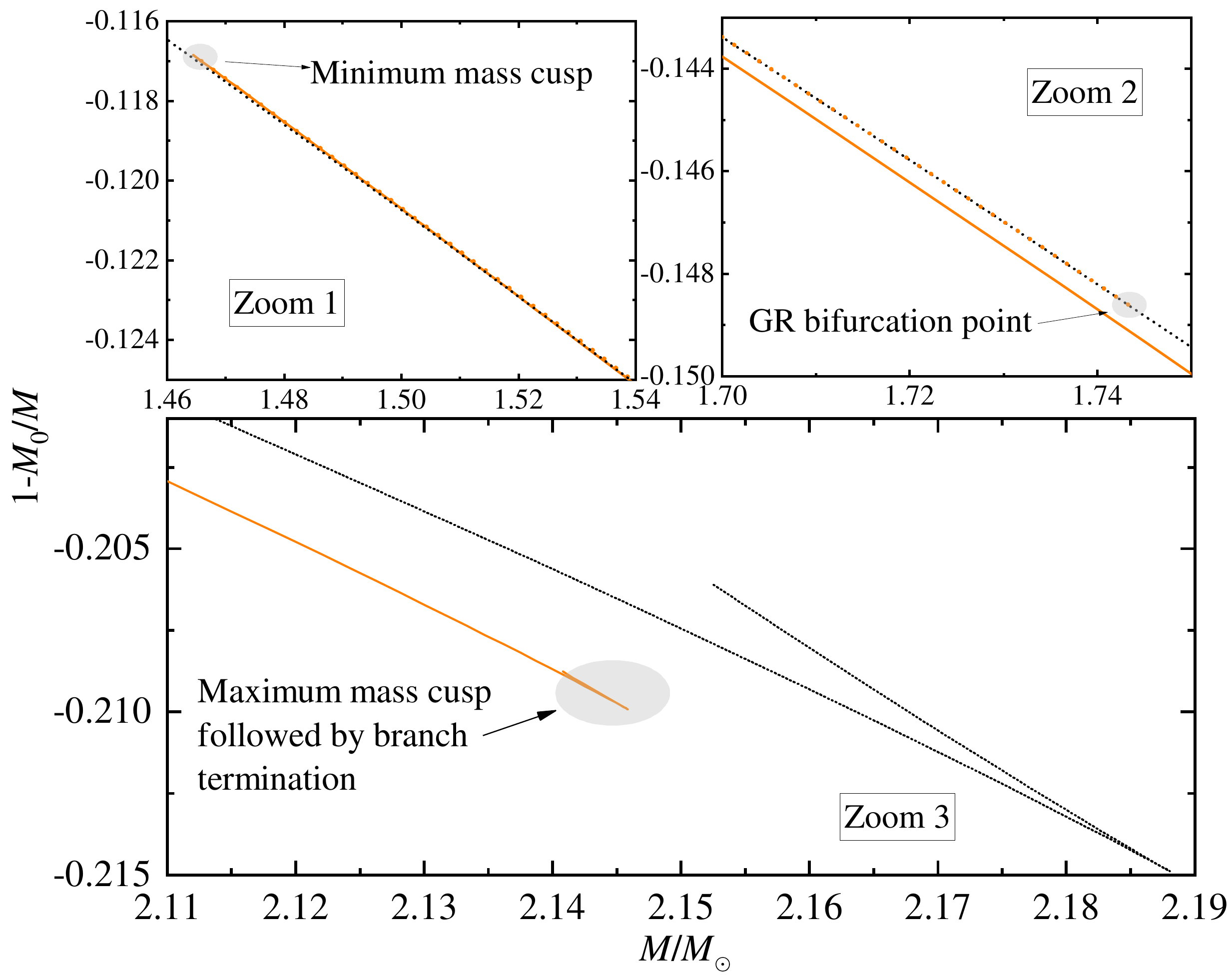}
	\caption{\textit{(left)} The binding energy of the star as a function of its mass for GR and a single combination $(\lambda,\beta,\kappa)$. \textit{(right:)} Zoomed in the minimal mass cusp, the bifurcation point, and the maximal mass cusp. The results are for coupling function $f_1(\varphi)$.}
	\label{fig:D_c1}
\end{figure}

In Fig. \ref{fig:D_c1}, we present the binding energy $\textrm{BE} = 1 - M_0/M$ as a function of the mass of the star. We plot sequences of solutions for the GR case and a single representative set of values $(\lambda,\beta,\kappa) = (10, 400, 1050)$ for which the two scalarized branches stitched at a minimal mass appear. The binding energy provides us with an indication that the going downwards branch is unstable. The reason is that a cusp, analogous to the one associated with the instability beyond the maximum mass model, can be observed at the minimal masses where the two scalarized branches meet. Such a cusp is normally associated with a change of stability. Since the binding energy of the small branch between the bifurcation point and the minimum mass is smaller (by absolute value) than the other scalarized branch, it should be the unstable one. The scalarized branch after the minimum mass point might be stable since in most of its extent it has larger binding energy (by absolute value) than GR. Of course,  if it reaches a maximum mass, another cusp forms and the scalarized branch loses stability again. For completeness, we also plot the GR solutions beyond the maximum mass model that are unstable. They correspond to the second black line in the lower right corner of the plot. In the right panel, we present the three most interesting parts of the scalarized branch zoomed in -- the minimal mass cusp (where most probably the unstable scalarized branch transforms into a stable one), the bifurcation point, and the maximal mass cusp (where the scalarized branch most probably loses stability again).

\subsection{Coupling function $f_2(\varphi)$}

We proceed our study with the coupling function $f_2(\varphi)$. In this case we employ $(\lambda,\beta) = (20,800)$ and $(\lambda,\beta) = (40,3200)$. Just like in the $f_1$ case, the parameters are chosen so that for one of the $(\lambda,\beta)$ pairs the bifurcation point is located at a relatively small neutron star mass while for the other choice of $(\lambda,\beta)$ scalarized solutions appear close to the maximum neutron star mass.

\begin{figure}[tbp]
	\includegraphics[width=0.45\textwidth]{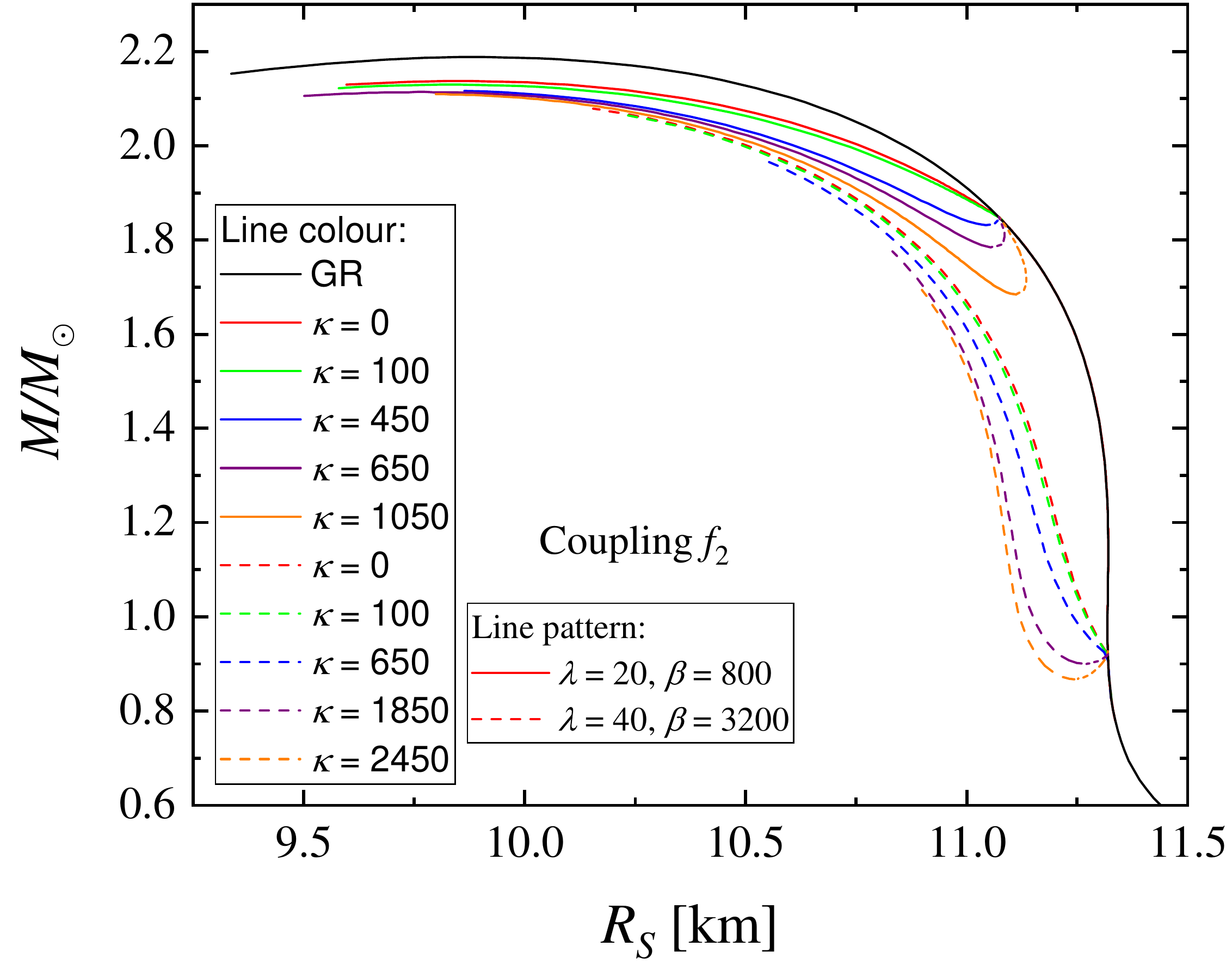}
	\includegraphics[width=0.45\textwidth]{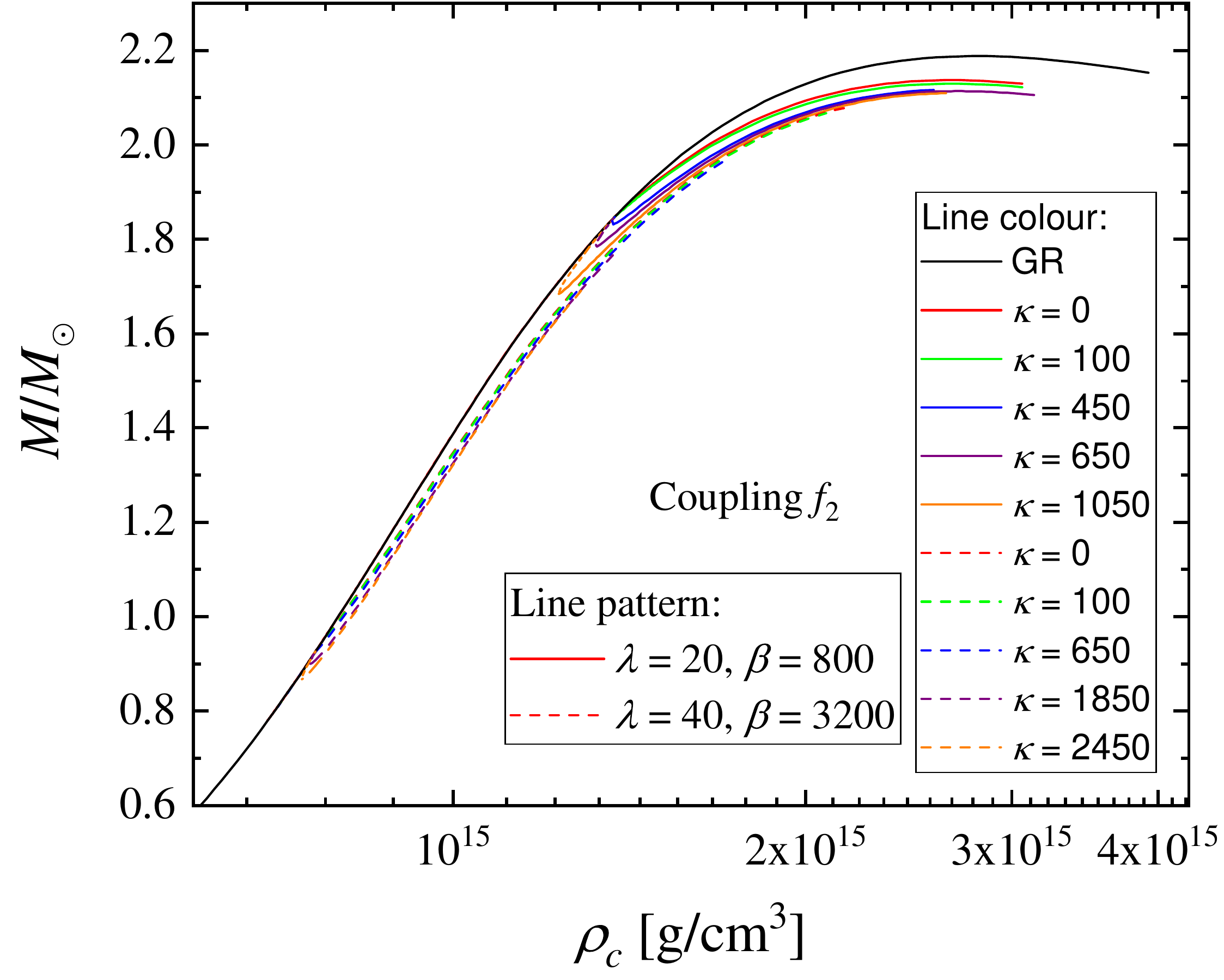}
        \includegraphics[width=0.45\textwidth]{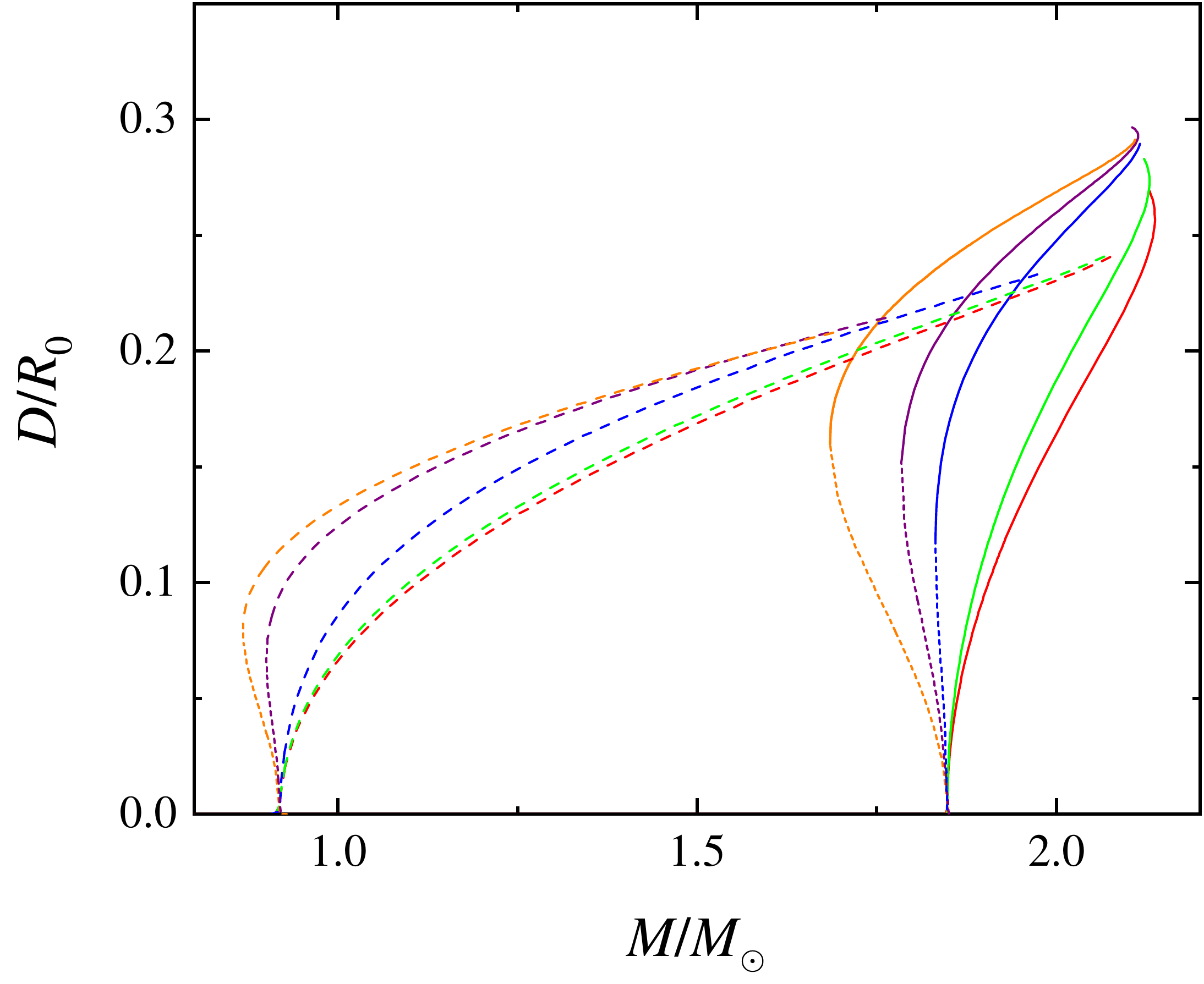}
	\caption{\textit{(top, left)} Neutron star mass as a function of its radius for several choices of the coupling parameters $\lambda,\beta$ and $\kappa$. The different pairs $(\lambda,\beta)$ are depicted with different line patterns, while different values of $\kappa$ are shown in different colours. \textit{(top, right)} Neutron star mass as a function of its central energy density for the same sequences of solution. \textit{(bottom)} The scalar charge, in units of $R_0$, as a function of the mass of the star. The presented results in these graphs are for the coupling function $f_2(\varphi)$.}
	\label{fig:M_c2}
\end{figure}

In the top panels of Fig. \ref{fig:M_c2}, we plot the mass of the neutron star as a function of its radius (left), and the mass of the star as a function of the central energy density (right). The same transitioning between spontaneous and nonlinear scalarization as in the $f_1$ case can be observed here when $\kappa$ is varied. For low values of $\kappa$ the scalarized branch resembles the standard sGB branch, and with increasing $\kappa$ a branch which goes downwards from the bifurcation point appears. After reaching a minimal mass, this branch transitions smoothly into second scalarized branch, going upwards. Hence, similarly to the $f_1$ case, nonlinear scalarization occurs for a wide range of parameters and central energy densities. Thus, gravitational phrase transition is possible. A major difference between the two cases, though, is that for $f_2$, the transition between the two types of behaviours happens only for significantly higher values of $\kappa$ compared to the  $f_1$ case, in particular when the energy density at which the bifurcation occurs is low. In the bottom panel of Fig. \ref{fig:M_c2}, for completeness, we plot the scalar charge, in units of $R_0$, as a function of the mass of the star. Interestingly, contrary to the $f_1$ coupling function, as well as the DEF scalarized neutron stars, it always increases never reaching a maximum.

\begin{figure}[btp]
	\includegraphics[width=0.49\textwidth]{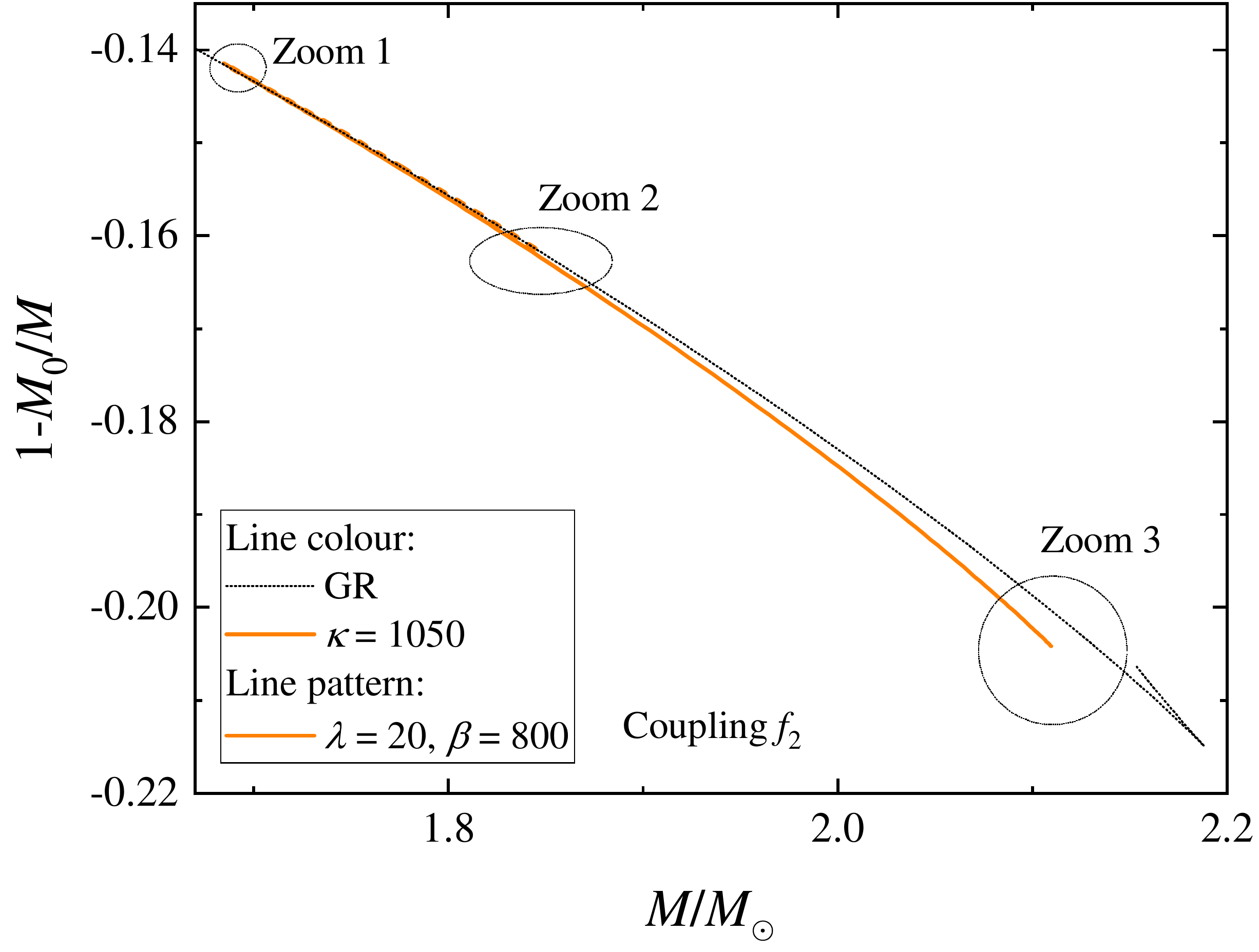}
	\includegraphics[width=0.49\textwidth]{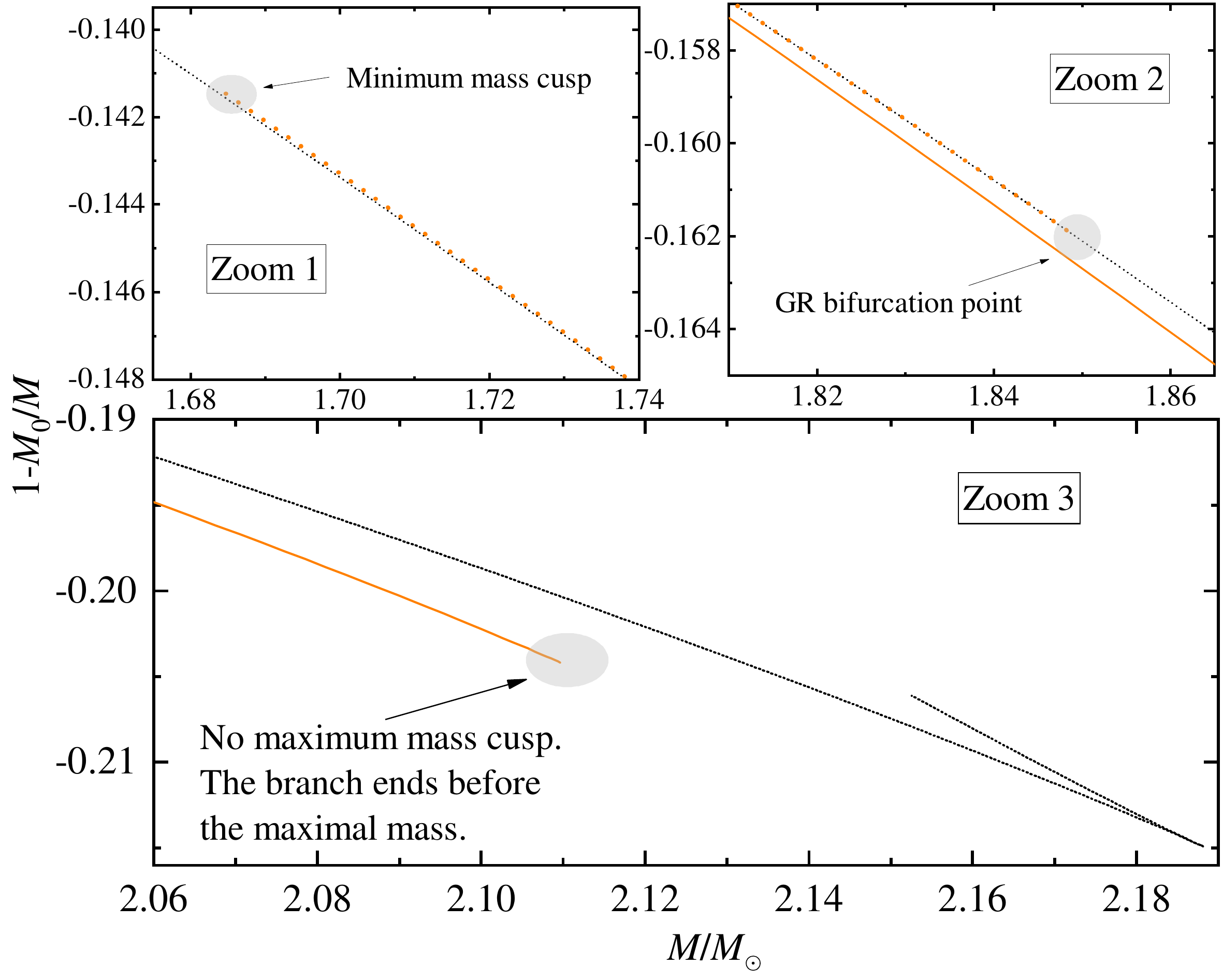}
	\caption{ Left: The binding energy of the star as a function of the mass of the star for GR and a single combination $(\lambda,\beta,\kappa)$. Right: Zoomed in the minimal mass cusp, the bifurcation point and the termination point for the branch (for the given set of parameters, the branch is terminated before the maximal mass, hence no maximal mass cusp). The results are for coupling function $f_2(\varphi)$. }
	\label{fig:D_c2}
\end{figure}

In Fig. \ref{fig:D_c2}, left panel, we present the binding energy as a function of the stellar mass for GR and a single set of values $(\lambda, \beta, \kappa) = (20, 800, 1050)$ for which the two branches and a minimal mass appear. The observed behaviour is the same as for the coupling function $f_1$. Two cusps exist -- one at the maximal mass for the GR sequence where the neutron stars loose linear stability (not present for the particular sGB branch since it is terminated before the maximum of the mass), and one at the minimal mass of the scalarized branch where a change of stability happens as well. Thus, the branch between the bifurcation point and the minimum mass is the unstable one. In the right panel we plot zoomed in the minimal mass cusp, the bifurcation point and the termination point of the scalarized branch. 

\subsection{Stability analysis through nonlinear saturation simulations}
\label{sec:nonlinear_saturation}

As explained above, given a specific central energy density $\rho_c$, the elliptic equations describing equilibrium neutron stars may have different number of solutions depending on the coupling parameters $\lambda$, $\beta$, and $\kappa$ (and also on the coupling function $f(\varphi)$). Besides the GR neutron stars, there may exist one or even two scalarized solutions with the same mass $M$ but slightly smaller radius $R$ compared to the $\varphi=0$ case; above, we have argued about their stability based on the presence of bifurcation points in the binding energy-mass diagrams, namely Figs. \ref{fig:D_c1} and \ref{fig:D_c2}.

In this section, we provide yet another argument by employing nonlinear time evolution simulations of perturbations that the conjectures about (in)stability given above, indeed hold. We consider some initial scalar field $\varphi(t, r)$ on a static (pure GR) background solution and study its evolution in time; the evolution of such a scalar field will be governed by Eq.~\eqref{eq:fe_scalar}. In this section only, we employ isotropic coordinates\footnote{Note that we use the same identifiers $t$, $r$, $\theta$, and $\phi$ for the isotropic coordinates as we did in the Schwarzschild-type metric, Eq.~\eqref{eq:metric}; however, as we use isotropic coordinates exclusively in this subsection, no confusion should arise.} in which
\begin{align}
    ds^2= - e^{2\nu(r)}dt^2 + e^{2\psi(r)} \left( dr^2 + r^2 d\theta^2 + \sin^2\theta d\phi^2 \right).
\end{align}   
Using these coordinates, the wave equation for $\varphi$ resulting from Eq.~\eqref{eq:fe_scalar} reads
\begin{align}
    e^{2\psi - 2\nu} \frac{\partial^2 \varphi}{\partial t^2}
        & = \frac{\partial^2 \varphi}{\partial r^2}
            + \frac{1}{r^2} \frac{\partial^2 \varphi}{\partial\theta^2}
            + \left( \nu_{,r} + \psi_{,r} + \frac{2}{r}\right) \frac{\partial \varphi}{\partial r}
            + \frac{1}{r^2} \left( \nu_{,\theta} + \psi_{,\theta} + \cot\theta \right) \frac{\partial \varphi}{\partial \theta}
    \nonumber\\
        & \qquad
            + e^{2\psi} \frac{\lambda^2}{4} \frac{d f(\varphi)}{d \varphi} R_{GB}^2,
\end{align}
where partial derivatives of the background quantities are denoted in the usual comma-notation, no $\phi$-derivatives appear since we consider only axisymmetric (i.e., $m=0$) scalar fields, and the Gauss-Bonnet invariant in isotropic coordinates is given by
\begin{align}
    R_{GB}^2
        & = 8 e^{-4\psi} \left[
            \left(\frac{2}{r}  + \psi_{,r}\right) \psi_{,r} \left( \nu_{,rr} + \nu_{,r}^2 \right)
            + \left( \frac{2}{r^2}  - \psi_{,r}^2 \right) \psi_{,r} \nu_{,r}
            + 2 \left( \frac{1}{r} + \psi_{,r} \right) \psi_{,rr} \nu_{,r}
            \right].
\end{align}
In contrast to perturbative studies in which quadratic or higher order terms of some quantity are neglected, here, we keep nonlinear terms in $\varphi$; in the wave equation, these appear in the last term due to $df(\varphi) / d\varphi$ which we do not treat perturbatively. Instead, we plug in the derivative of the chosen coupling function $f(\varphi)$ which results in nonlinear terms in the case that the coupling function contains terms of higher than quadratic order in $\varphi$. Note that the coefficients of this differential equation contain quantities of a static (pure GR) solution (i.\,e., $\nu$, $\psi$ and $R_{GB}^2$); these do not change during the simulation because we are working in the decoupling limit approximation. Namely, we neglect the backreaction of the scalar field on the metric and the fluid and thus their evolution is the same as in GR.

The nonlinear terms do not pose a numerical challenge regarding the stability of the time evolution, however, the time evolution drastically changes its characteristics: Given an initial scalar perturbation, the solution will also depend on the amplitude of the initial data. In the case where the $\varphi = 0$ GR solution is unstable (i.\,e., for $M \gtrsim 1.74\,M_\odot$), the scalar field will, for arbitrary small amplitude, grow exponentially and saturate at some limiting value $\varphi(r)$. When the $\varphi(r) = 0$ GR solution as well as a scalarised solution are linearly stable, the initial scalar field will, for large enough initial amplitudes, redistribute and converge to the non-zero scalarised solution. If its initial amplitude is smaller than a given threshold, it is radiated away entirely and the GR solution is recovered. When the scalar field converges to the scalarised solution, the limiting value is very similar (within a few percent), but not identical, to the exact solution $\varphi(r)$ of the scalarized equilibrium configuration; the discrepancy arises because the time evolution is performed on a static neutron star background, thus the scalar field backreaction is neglected.

\begin{figure}[tbph]
    \includegraphics[width=0.7\textwidth]{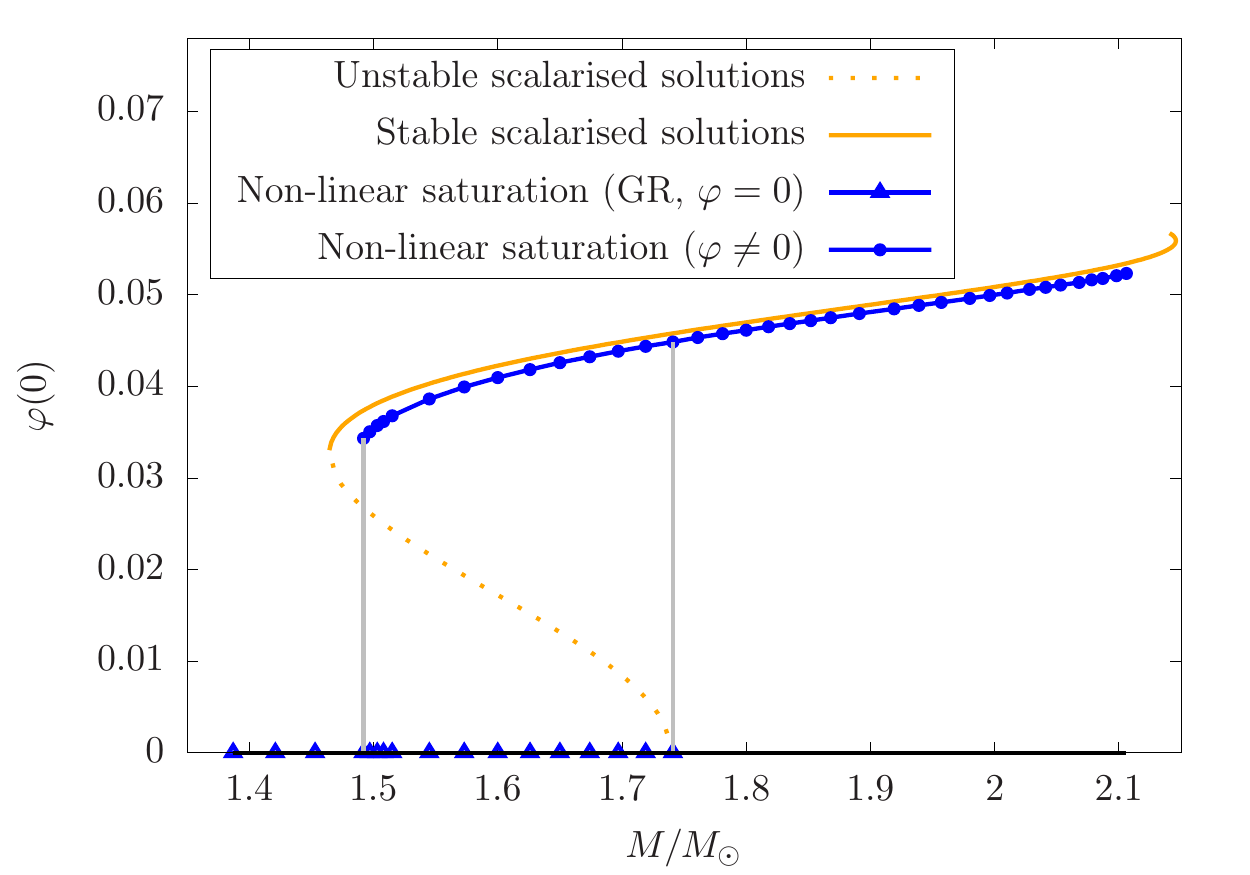}
    \caption{Comparison of the scalar field $\varphi(0)$ at the center of the star between the solution of the nonlinear elliptic problem (orange) and the saturated nonlinear time evolution (blue); the results here are shown for $(\lambda, \beta, \kappa) = (10, 400, 1050)$ using the coupling function $f_1(\varphi)$. We distinguish scalarized solutions (blue dots) and pure GR solutions (blue triangles). Where scalarized solutions exist, i.e., for $M > 1.49\,M_\odot$, the results of both approaches agree very well; additionally, the nonlinear time evolution strongly suggests that, within the mass range $M \in [1.49, 1.74]\,M_\odot$ where two scalarized solutions exist, only one of them is stable (the saturation simulation never converges to the intermediate solution). }
    \label{fig:phi0}
\end{figure}

Results for the nonlinear time-evolution are illustrated in Fig.~\ref{fig:phi0} for the case $(\lambda, \beta, \kappa) = (10, 400, 1050)$ using the coupling function $f_1(\varphi)$. For small central energy densities $\rho_c$ (for which the mass of the neutron star is $M \lesssim 1.49\,M_\odot$), only the $\varphi=0$ GR solution exists and any scalar perturbation exponentially decays, resulting in $\varphi(0) \rightarrow 0$ at the center of the star. For large $\rho_c$ (for which $M \gtrsim 1.74\,M_\odot$) the opposite happens. Even very small initial scalar perturbations will continue to grow and eventually saturate at some limiting scalar field $\varphi(r)$; we can confirm that this function as a whole deviates only negligibly from the ``correct'' scalarized solution of the same mass. In Fig.~\ref{fig:phi0}, this similarity is reflected by the fact that the blue and orange curves almost coincide. Last, we need to comment on the central densities $\rho_c$ for which $M \in [1.49, 1.74]\,M_\odot$, i.e., in between the two grey vertical lines; for those values, there are in fact three individual solutions: the $\varphi=0$ GR solution and two scalarized solutions, one of which is potentially unstable. We find that the amplitude of the initial scalar perturbation determines whether the scalar field will decay and eventually converge to the GR solution (small initial amplitude, shown with blue triangles on the $x$-axis) or if the scalar field grows and eventually converges to the stable scalarized solution (large initial amplitude, the upper branch of blue dots); however, the intermediate scalarized solutions (shown as an orange dotted line in the figure) are never realised as the final state of our time evolutions. We take this as further evidence that these solutions are indeed dynamically unstable. Close to the minimum mass of the scalarized branch, the saturated solutions (in blue) deviate a small bit from the scalarized (in orange) solutions and also the mass window in which we find two different solutions via the saturation simulation differs a bit from the mass window found via the solution of the elliptic equations; this is simply (again) because our time evolution runs on a static GR background with neglected scalar field backreaction.

\section{Discussion}

We consider equilibrium configurations of neutron stars in scalar-Gauss-Bonnet gravity with coupling functions possessing higher than quadratic order terms in the scalar field. In the same way that nonlinear scalarization of black holes takes place, we find that for such couplings similar phenomenon also occurs in neutron stars. Namely, for certain regions of the parameters space scalarized neutron star solutions co-exist with linearly stable GR solutions and the transition between the two can happen only with a large nonlinear perturbation.

We study two coupling functions that differ only by their sign but have an important distinction---only one of them allows for black hole scalarization. Those functions lead to a mixture of curvature-induced spontaneous scalarization (due to the quadratic scalar field term) and nonlinear scalarization (due to the quartic term). Once the quartic term starts to dominate, two potentially linearly stable branches coexist for a limited range of masses---one branch is the well-known GR branch, the other branch possesses a scalar field but exists only beyond some central energy density of the star that depends on the parameters of the theory. Those two branches can be connected only via a sequence of equilibrium scalarized configurations for which we find strong evidence that it is unstable. Thus, a transition between those two potentially stable branches may happen only discontinuously, a process that may resemble a first-order matter phase transition. In contrast to the Damour-Esposito-Farese  model, where similar properties of the solutions were recently discovered, the appearance of two potentially stable disconnected branches happens for a much larger range of theory parameters and central energy densities. In addition, the size of the ``gap'' between them can be easily controlled through the coupling parameters.

In order to check stability, we have also performed nonlinear perturbation evolution simulations where we evolve a small scalar perturbation on top of a GR non-rotating neutron star configurations without scalar field. We find that the pure GR solution is unstable (given that its central energy density is beyond the parameter-dependent bifurcation threshold) as any small scalar perturbation rapidly grows and converges against the scalarized solution (approximately, since we do not account for the backreaction of the scalar field on the metric and the fluid). In the regime where the scalarized phases co-exist with the linearly stable GR solutions, the scalarization can be ignited only by a strong scalar field perturbation. What is also important is that the time evolution code always converges to one of the scalarized branches speaking in favour of the conjecture that the other one is unstable. The conclusions based on the nonlinear evolution are in agreement with the analysis relying on the examination of the stellar binding energy and the presence of cusps.

The discontinuous transition between the two stable solutions could have significant astrophysical implications that deserve further study. They will mimic very well the astrophysical effects predicted for neutron stars exhibiting matter phase transitions but new phenomenology might exist due to the presence of a scalar field. It would be also interesting to study whether gravitational phase transitions in sGB gravity possess qualitative deviations from the DEF model. Further, the stability analysis of the solutions could be extended; our current claims are based on the binding energy, the analogy to the black hole case, and nonlinear evolution of perturbations. Thus, a proper linear analysis should be conducted.

 \section*{Acknowledgement}
This study is in part financed by the European Union-NextGenerationEU, through the National Recovery and Resilience Plan of the Republic of Bulgaria, project No. BG-RRP-2.004-0008-C01. DD acknowledges financial support via an Emmy Noether Research Group funded by the German Research Foundation (DFG) under grant no. DO 1771/1-1. The authors acknowledge support by the High Performance and Cloud Computing Group at the Zentrum für Datenverarbeitung of the University of Tübingen, the state of Baden-Württemberg through bwHPC and the German Research Foundation (DFG) through Grant No. INST 37/935-1 FUGG.

\bibliographystyle{unsrt}
\bibliography{references}

\begin{thebibliography}{10}

\bibitem{Berti:2015itd}
Emanuele Berti et~al.
\newblock {Testing General Relativity with Present and Future Astrophysical
  Observations}.
\newblock {\em Class. Quant. Grav.}, 32:243001, 2015.

\bibitem{Doneva:2017jop}
Daniela~D. Doneva and George Pappas.
\newblock {Universal Relations and Alternative Gravity Theories}.
\newblock {\em Astrophys. Space Sci. Libr.}, 457:737--806, 2018.

\bibitem{Shao:2019gjj}
Lijing Shao.
\newblock {Degeneracy in Studying the Supranuclear Equation of State and
  Modified Gravity with Neutron Stars}.
\newblock {\em AIP Conf. Proc.}, 2127(1):020016, 2019.

\bibitem{LIGOScientific:2018dkp}
B.~P. Abbott et~al.
\newblock {Tests of General Relativity with GW170817}.
\newblock {\em Phys. Rev. Lett.}, 123(1):011102, 2019.

\bibitem{Zhao:2022vig}
Junjie Zhao, Paulo C.~C. Freire, Michael Kramer, Lijing Shao, and Norbert Wex.
\newblock {Closing a spontaneous-scalarization window with binary pulsars}.
\newblock {\em Class. Quant. Grav.}, 39(11):11LT01, 2022.

\bibitem{Silva:2020acr}
Hector~O. Silva, A.~Miguel Holgado, Alejandro C\'ardenas-Avenda\~no, and
  Nicol\'as Yunes.
\newblock {Astrophysical and theoretical physics implications from
  multimessenger neutron star observations}.
\newblock {\em Phys. Rev. Lett.}, 126(18):181101, 2021.

\bibitem{Kuan:2022oxs}
Hao-Jui Kuan, Arthur~G. Suvorov, Daniela~D. Doneva, and Stoytcho~S. Yazadjiev.
\newblock {Gravitational Waves from Accretion-Induced Descalarization in
  Massive Scalar-Tensor Theory}.
\newblock {\em Phys. Rev. Lett.}, 129(12):121104, 2022.

\bibitem{kam81}
Burkhard Kampfer.
\newblock {On the Possibility of Stable Quark and Pion Condensed Stars}.
\newblock {\em J. Phys. A}, 14:L471--L475, 1981.

\bibitem{Glendenning:1998ag}
Norman~K. Glendenning and Christiane Kettner.
\newblock {Nonidentical neutron star twins}.
\newblock {\em Astron. Astrophys.}, 353:L9, 2000.

\bibitem{sch00}
K.~Schertler, C.~Greiner, J.~Schaffner-Bielich, and M.~H. Thoma.
\newblock {Quark phases in neutron stars and a 'third family' of compact stars
  as a signature for phase transitions}.
\newblock {\em Nucl. Phys. A}, 677:463--490, 2000.

\bibitem{shaf02}
Jurgen Schaffner-Bielich, Matthias Hanauske, Horst Stoecker, and Walter
  Greiner.
\newblock {Phase transition to hyperon matter in neutron stars}.
\newblock {\em Phys. Rev. Lett.}, 89:171101, 2002.

\bibitem{Most:2018eaw}
Elias~R. Most, L.~Jens Papenfort, Veronica Dexheimer, Matthias Hanauske, Stefan
  Schramm, Horst St\"ocker, and Luciano Rezzolla.
\newblock {Signatures of quark-hadron phase transitions in general-relativistic
  neutron-star mergers}.
\newblock {\em Phys. Rev. Lett.}, 122(6):061101, 2019.

\bibitem{Bauswein:2018bma}
Andreas Bauswein, Niels-Uwe~F. Bastian, David~B. Blaschke, Katerina
  Chatziioannou, James~A. Clark, Tobias Fischer, and Micaela Oertel.
\newblock {Identifying a first-order phase transition in neutron star mergers
  through gravitational waves}.
\newblock {\em Phys. Rev. Lett.}, 122(6):061102, 2019.

\bibitem{Weih:2019xvw}
Lukas~R. Weih, Matthias Hanauske, and Luciano Rezzolla.
\newblock {Postmerger Gravitational-Wave Signatures of Phase Transitions in
  Binary Mergers}.
\newblock {\em Phys. Rev. Lett.}, 124(17):171103, 2020.

\bibitem{Liebling:2020dhf}
Steven~L. Liebling, Carlos Palenzuela, and Luis Lehner.
\newblock {Effects of High Density Phase Transitions on Neutron Star Dynamics}.
\newblock {\em Class. Quant. Grav.}, 38(11):115007, 2021.

\bibitem{Most:2019onn}
Elias~R. Most, L.~Jens~Papenfort, Veronica Dexheimer, Matthias Hanauske, Horst
  Stoecker, and Luciano Rezzolla.
\newblock {On the deconfinement phase transition in neutron-star mergers}.
\newblock {\em Eur. Phys. J. A}, 56(2):59, 2020.

\bibitem{Blacker:2020nlq}
Sebastian Blacker, Niels-Uwe~F. Bastian, Andreas Bauswein, David~B. Blaschke,
  Tobias Fischer, Micaela Oertel, Theodoros Soultanis, and Stefan Typel.
\newblock {Constraining the onset density of the hadron-quark phase transition
  with gravitational-wave observations}.
\newblock {\em Phys. Rev. D}, 102(12):123023, 2020.

\bibitem{Blaschke:2019tbh}
David Blaschke, David~Edwin Alvarez-Castillo, Alexander Ayriyan, Hovik
  Grigorian, Noshad~Khosravi Largani, and Fridolin Weber.
\newblock {\em {Astrophysical aspects of general relativistic mass twin
  stars}}, pages 207--256.
\newblock 2020.

\bibitem{dam93}
Thibault Damour and Gilles Esposito-Farese.
\newblock {Nonperturbative strong field effects in tensor - scalar theories of
  gravitation}.
\newblock {\em Phys. Rev. Lett.}, 70:2220--2223, 1993.

\bibitem{Ramazanoglu16}
Fethi~M. Ramazano\u{g}lu and Frans Pretorius.
\newblock {Spontaneous Scalarization with Massive Fields}.
\newblock {\em Phys. Rev. D}, 93(6):064005, 2016.

\bibitem{Yazadjiev16}
Stoytcho~S. Yazadjiev, Daniela~D. Doneva, and Dimitar Popchev.
\newblock {Slowly rotating neutron stars in scalar-tensor theories with a
  massive scalar field}.
\newblock {\em Phys. Rev. D}, 93(8):084038, 2016.

\bibitem{Christian:2019qer}
Jan-Erik Christian and J\"urgen Schaffner-Bielich.
\newblock {Twin Stars and the Stiffness of the Nuclear Equation of State:
  Ruling Out Strong Phase Transitions below $1.7n_0$ with the New NICER Radius
  Measurements}.
\newblock {\em Astrophys. J. Lett.}, 894(1):L8, 2020.

\bibitem{Benic:2014jia}
Sanjin Benic, David Blaschke, David~E. Alvarez-Castillo, Tobias Fischer, and
  Stefan Typel.
\newblock {A new quark-hadron hybrid equation of state for astrophysics - I.
  High-mass twin compact stars}.
\newblock {\em Astron. Astrophys.}, 577:A40, 2015.

\bibitem{Silva:2017uqg}
Hector~O. Silva, Jeremy Sakstein, Leonardo Gualtieri, Thomas~P. Sotiriou, and
  Emanuele Berti.
\newblock {Spontaneous scalarization of black holes and compact stars from a
  Gauss-Bonnet coupling}.
\newblock {\em Phys. Rev. Lett.}, 120(13):131104, 2018.

\bibitem{Doneva:2017duq}
Daniela~D. Doneva and Stoytcho~S. Yazadjiev.
\newblock {Neutron star solutions with curvature induced scalarization in the
  extended Gauss-Bonnet scalar-tensor theories}.
\newblock {\em JCAP}, 04:011, 2018.

\bibitem{Xu:2021kfh}
Rui Xu, Yong Gao, and Lijing Shao.
\newblock {Neutron stars in massive scalar-Gauss-Bonnet gravity: Spherical
  structure and time-independent perturbations}.
\newblock {\em Phys. Rev. D}, 105(2):024003, 2022.

\bibitem{Doneva:2021tvn}
Daniela~D. Doneva and Stoytcho~S. Yazadjiev.
\newblock {Beyond the spontaneous scalarization: New fully nonlinear mechanism
  for the formation of scalarized black holes and its dynamical development}.
\newblock {\em Phys. Rev. D}, 105(4):L041502, 2022.

\bibitem{Doneva:2022byd}
Daniela~D. Doneva, Alex Va\~n\'o Vi\~nuales, and Stoytcho~S. Yazadjiev.
\newblock {Dynamical descalarization with a jump during a black hole merger}.
\newblock {\em Phys. Rev. D}, 106(6):L061502, 2022.

\bibitem{Akmal:1998cf}
A.~Akmal, V.~R. Pandharipande, and D.~G. Ravenhall.
\newblock {The Equation of state of nucleon matter and neutron star structure}.
\newblock {\em Phys. Rev. C}, 58:1804--1828, 1998.

\bibitem{Read:2008iy}
Jocelyn~S. Read, Benjamin~D. Lackey, Benjamin~J. Owen, and John~L. Friedman.
\newblock {Constraints on a phenomenologically parameterized neutron-star
  equation of state}.
\newblock {\em Phys. Rev. D}, 79:124032, 2009.

\bibitem{Damour:1993hw}
Thibault Damour and Gilles Esposito-Farese.
\newblock {Nonperturbative strong field effects in tensor - scalar theories of
  gravitation}.
\newblock {\em Phys. Rev. Lett.}, 70:2220--2223, 1993.

\bibitem{Stefanov:2007eq}
Ivan~Zh. Stefanov, Stoytcho~S. Yazadjiev, and Michail~D. Todorov.
\newblock {Phases of 4D scalar-tensor black holes coupled to Born-Infeld
  nonlinear electrodynamics}.
\newblock {\em Mod. Phys. Lett. A}, 23:2915--2931, 2008.

\bibitem{Doneva:2017bvd}
Daniela~D. Doneva and Stoytcho~S. Yazadjiev.
\newblock {New Gauss-Bonnet Black Holes with Curvature-Induced Scalarization in
  Extended Scalar-Tensor Theories}.
\newblock {\em Phys. Rev. Lett.}, 120(13):131103, 2018.

\bibitem{Antoniou:2017acq}
G.~Antoniou, A.~Bakopoulos, and P.~Kanti.
\newblock {Evasion of No-Hair Theorems and Novel Black-Hole Solutions in
  Gauss-Bonnet Theories}.
\newblock {\em Phys. Rev. Lett.}, 120(13):131102, 2018.

\bibitem{Doneva:2022ewd}
Daniela~D. Doneva, Fethi~M. Ramazano\u{g}lu, Hector~O. Silva, Thomas~P.
  Sotiriou, and Stoytcho~S. Yazadjiev.
\newblock {Scalarization}.
\newblock 11 2022.

\bibitem{Blazquez-Salcedo:2022omw}
Jose~Luis Bl\'azquez-Salcedo, Daniela~D. Doneva, Jutta Kunz, and Stoytcho~S.
  Yazadjiev.
\newblock {Radial perturbations of scalar-Gauss-Bonnet black holes beyond
  spontaneous scalarization}.
\newblock {\em Phys. Rev. D}, 105(12):124005, 2022.

\bibitem{Zhang:2023jei}
Shao-Jun Zhang.
\newblock {Nonlinear instability and scalar clouds of spherical exotic compact
  objects in scalar-Gauss-Bonnet theory}.
\newblock 4 2023.

\bibitem{Lai:2023gwe}
Meng-Yun Lai, De-Cheng Zou, Rui-Hong Yue, and Yun~Soo Myung.
\newblock {Nonlinearly scalarized rotating black holes in
  Einstein-scalar-Gauss-Bonnet theory}.
\newblock 4 2023.

\bibitem{Jiang:2023yyn}
Jia-Yan Jiang, Qian Chen, Yunqi Liu, Yu~Tian, Wei Xiong, Cheng-Yong Zhang, and
  Bin Wang.
\newblock {Type I critical dynamical scalarization and descalarization in
  Einstein-Maxwell-scalar theory}.
\newblock 6 2023.

\bibitem{Blazquez-Salcedo:2020nhs}
Jose~Luis Bl\'azquez-Salcedo, Carlos A.~R. Herdeiro, Jutta Kunz, Alexandre~M.
  Pombo, and Eugen Radu.
\newblock {Einstein-Maxwell-scalar black holes: the hot, the cold and the
  bald}.
\newblock {\em Phys. Lett. B}, 806:135493, 2020.

\bibitem{LuisBlazquez-Salcedo:2020rqp}
Jose Luis Bl\'azquez-Salcedo, Carlos A.~R. Herdeiro, Sarah Kahlen, Jutta Kunz,
  Alexandre~M. Pombo, and Eugen Radu.
\newblock {Quasinormal modes of hot, cold and bald
  Einstein\textendash{}Maxwell-scalar black holes}.
\newblock {\em Eur. Phys. J. C}, 81(2):155, 2021.

\bibitem{Silva:2018qhn}
Hector~O. Silva, Caio F.~B. Macedo, Thomas~P. Sotiriou, Leonardo Gualtieri,
  Jeremy Sakstein, and Emanuele Berti.
\newblock {Stability of scalarized black hole solutions in scalar-Gauss-Bonnet
  gravity}.
\newblock {\em Phys. Rev. D}, 99(6):064011, 2019.

\bibitem{Minamitsuji:2018xde}
Masato Minamitsuji and Taishi Ikeda.
\newblock {Scalarized black holes in the presence of the coupling to
  Gauss-Bonnet gravity}.
\newblock {\em Phys. Rev. D}, 99(4):044017, 2019.

\bibitem{Danchev:2021tew}
Victor~I. Danchev, Daniela~D. Doneva, and Stoytcho~S. Yazadjiev.
\newblock {Constraining scalarization in scalar-Gauss-Bonnet gravity through
  binary pulsars}.
\newblock {\em Phys. Rev. D}, 106(12):124001, 2022.

\bibitem{Kuan:2023trn}
Hao-Jui Kuan, Alan Tsz-Lok Lam, Daniela~D. Doneva, Stoytcho~S. Yazadjiev,
  Masaru Shibata, and Kenta Kiuchi.
\newblock {Dynamical Scalarization during Neutron Star Mergers in
  scalar-Gauss-Bonnet Theory}.
\newblock 2 2023.

\bibitem{Sperhake:2017itk}
Ulrich Sperhake, Christopher~J. Moore, Roxana Rosca, Michalis Agathos, Davide
  Gerosa, and Christian~D. Ott.
\newblock {Long-lived inverse chirp signals from core collapse in massive
  scalar-tensor gravity}.
\newblock {\em Phys. Rev. Lett.}, 119(20):201103, 2017.

\end{thebibliography}
\end{document}